\documentclass{elsart}

\begin{document}

\begin{frontmatter}
\bibliographystyle{elsart-num}

\title{Reconstructing weak values without weak measurements}

\author{Lars M. Johansen}

\address{Department of Technology, Buskerud University College,
N-3601 Kongsberg, Norway}

\ead{lars.m.johansen@hibu.no}

\date{\today}

\begin{abstract}

I propose a scheme for reconstructing the weak value of an
observable without the need for weak measurements. The
post-selection in weak measurements is replaced by an initial
projector measurement. The observable can be measured using any form
of interaction, including projective measurements. The
reconstruction is effected by measuring the change in the
expectation value of the observable due to the projector
measurement. The weak value may take nonclassical values if the
projector measurement disturbs the expectation value of the
observable.

\end{abstract}


\end{frontmatter}

\section{Introduction}

A weak measurement is achieved either by using a measurement probe
with a large uncertainty
\cite{Aharonov+AlbertETAL-ResuMeasCompSpin:88} or by employing a
weak measurement interaction \cite{Johansen-WeakMeaswithArbi:04}. By
weakening the interaction, one may obtain an arbitrarily small
perturbation of the system state. Aharonov, Albert and Vaidman (AAV)
\cite{Aharonov+AlbertETAL-ResuMeasCompSpin:88} considered an
experimental scheme where the weak measurement is followed by a
projector measurement, usually called a post-selection. The average
of a weak measurement of an observable $\hat{B}$ conditioned on a
post-selection represented by a projector $\hat{P}_a^2=\hat{P}_a$
can be expressed as the real part of a complex ``weak value''
\cite{Aharonov+AlbertETAL-ResuMeasCompSpin:88,%
Johansen+Luis-NoncWeakMeas:04}
\begin{equation}
    \label{eq:def}
    B_w(a) = {\mathrm{Tr} \hat{\rho} \hat{P}_a \hat{B} \over
    \mathrm{Tr} \hat{\rho} \hat{P}_a}.
\end{equation}
AAV came to the surprising conclusion that this average may take
values outside the eigenvalue range. The theory predicts effects
such as a negative kinetic energy
\cite{Aharonov+PopescuETAL-MeasErroNegaKine:93}, negative photon
number \cite{Johansen+Luis-NoncWeakMeas:04} and negative
probabilities \cite{Molmer-Counstatweakmeas:01,%
Aharonov+BoteroETAL-ReviHardpara:02,%
Resch+LundeenETAL-Experealquanprob:04}
for suitably chosen sub-ensembles. The theory of weak measurements
has been applied in a number of areas (for a review see Ref.
\cite{Aharonov+Botero-QuanAverWeakValu:05}), and it has been
confirmed in experiments involving classical intense laser beams
\cite{Ritchie+StoryETAL-RealMeasWeakValu:91,%
Parks+CullinETAL-ObsemeasoptiAhar:98},
anomalous pulse propagation
\cite{Rohrlich+Aharonov-Cherradisupepart:02,%
Solli+McCormickETAL-FastLighSlowLigh:04,%
Brunner+AcinETAL-OptiTeleNetwWeak:03},
the quantum box problem
\cite{Aharonov+Vaidman-Compdescquansyst:91,%
Resch+LundeenETAL-Experealquanprob:04}
and single photons
\cite{Pryde+OBrienETAL-MeasQuanWeakValu:05,%
Wang+SunETAL-Expedemomethreal:06}.

Weak values are generally considered to be an artifact of weak
measurements with post-selection. Steps towards generalizing the
application of weak values to more general types of measurements
have been made in Refs. \cite{Aharonov+Botero-QuanAverWeakValu:05}
and \cite{Tollaksen+Aharonov-Non-WeakMeas:06}. However, it is not
known whether weak values can have an operational significance e.g.
in projective measurements. In this Letter, I will demonstrate that
weak values may be reconstructed from an initial projector
measurement followed by a measurement of the observable. The
measurement of the observable may take any form as long as it
reproduces expectation values. It could be e.g. a projective
measurement or a weak measurement. The projector measurement
disturbs the subsequent measurement of the observable. The
reconstruction relies on measuring the change in the expectation
value of the observable due to the projector measurement. The
imaginary part of the weak value is reconstructed by measuring this
change relative to a state that has been subjected to a selective
phase rotation of $\pi/2$.

\section{Reconstruction of weak values}

In weak measurements with post-selection, the post-selection
operation is represented by a projection operator $\hat{P}_a^2 =
\hat{P}_a$. In this Letter we propose a measurement scheme where
this projector measurement is performed first. A projective
measurement of this projector is a binary experiment with only two
possible outcomes: yes or no. According to the von
Neumann-L{\"u}ders projection postulate
\cite{Neumann-MathFounQuanMech:55,Lueders-UEbeZust:51}, the result
of a selective projector measurement is
\begin{equation}
    \label{eq:selective}
    \hat{\rho}_a^s = {\hat{P}_a \hat{\rho} \hat{P}_a \over
    \mathrm{Tr} \hat{\rho} \hat{P}_a}.
\end{equation}
This state is represented by a sub-ensemble of the initially
prepared ensemble. For a rank-one projector $\hat{P}_a = | a
\rangle\ \langle a |$ the selective state is independent of the
initial preparation $\hat{\rho}$ and equals $\hat{\rho}_a^s=| a
\rangle\ \langle a |$.

If the outcome of the projector measurement is disregarded, the
state is also changed. This nonselective state is
\cite{Neumann-MathFounQuanMech:55,Lueders-UEbeZust:51},
\begin{equation}
    \label{eq:nonselective}
    \hat{\rho}_a^n = \hat{P}_a \hat{\rho} \hat{P}_a +
    (1-\hat{P}_a )\hat{\rho} (1-\hat{P}_a).
\end{equation}
This state is represented by the complete initially prepared
ensemble.

We now calculate the expectation value of the observable $\hat{B}$
on the nonselective state $\hat{\rho}_a^n$. We expand the r.h.s. of
Eq. (\ref{eq:nonselective}), multiply both sides by $\hat{B}$ and
take the trace of both sides. After division by $\mathrm{Tr}
\hat{\rho} \hat{P}_a$ and rearranging of terms we arrive at the
expression
\begin{equation}
    \label{eq:realweak}
    \mathrm{Re} B_w(a)  = \mathrm{Tr} \hat{\rho}_a^s \hat{B} +
    {\mathrm{Tr} \hat{\rho} \hat{B} - \mathrm{Tr} \hat{\rho}_a^n
    \hat{B} \over 2 \, \mathrm{Tr} \hat{\rho} \hat{P}_a}.
\end{equation}
This expresses the real part of the weak value in terms of the
expectation value of $\hat{B}$ on the selective state
$\hat{\rho}_a^s$ plus a term which is proportional to the change in
the expectation value of the observable $\hat{B}$ due to the initial
nonselective projector measurement.

The first term on the r.h.s. is bounded by the eigenvalue spectrum.
Therefore, the weak value can only exceed the eigenvalue spectrum if
the last term on the r.h.s. is non-vanishing, which requires the
expectation value of the observable $\hat{B}$ to be different on the
original state $\hat{\rho}$ than on the nonselective state
$\hat{\rho}_a^n$.

All terms on the r.h.s. can be measured directly. Measurements must
be performed on two identically prepared sub-ensembles. One
sub-ensemble is subjected only to a measurement of $\hat{B}$. The
other sub-ensemble is subjected to a measurement of $\hat{P}_a$
followed by a measurement of $\hat{B}$. The measurements of
$\hat{B}$ may be of any form. It could e.g. be a projective
measurement or a weak measurement.

A case of particular interest is when also the observable $\hat{B}$
is a projector, $\hat{B} = \hat{P}_b$. In this case, the first term
on the r.h.s. of Eq. (\ref{eq:realweak}) is a probability between 0
and 1. The ``weak probability'' \cite{Molmer-Counstatweakmeas:01,%
Aharonov+BoteroETAL-ReviHardpara:02,%
Resch+LundeenETAL-Experealquanprob:04}
on the l.h.s. may exceed the classical range only if an intervening
measurement of $\hat{P}_a$ disturbs the probability of $\hat{P}_b$.
This gives an intuitively pleasing picture of the connection between
extended quasi-probabilities in quantum mechanics and measurement
disturbance. An early attempt at relating extended probabilities to
measurement disturbance was made by Prugove\v{c}ki
\cite{Prugovecki-TheoMeasIncoObse:67}.

In order to be able to reconstruct also the imaginary part of the
weak value, we introduce the unitary operator
\begin{equation}
    \hat{R}_a^{\phi} = 1+(e^{i \phi} -1) \hat{P}_a.
\end{equation}
It is easily checked that ${R}_a^{\phi} \hat{P}_a = e^{i \phi}
\hat{P}_a$, whereas ${R}_a^{\phi} \hat{P}_b = \hat{P}_b$ for any
projector $\hat{P}_b$ orthogonal to $\hat{P}_a$. The operator
${R}_a^{\phi}$ imparts a phase change only on the projector
$\hat{P}_a$, but does not change any other orthogonal projectors. We
shall therefore refer to $\hat{P}_a$ as a selective phase rotation
operator. It can be implemented at time $t_0$ e.g. by adding to the
Hamiltonian a term $- \phi \delta(t-t_0) \hat{P}_a$. The state after
an arbitrary selective phase rotation is $\hat{\rho}_a^\phi =
\hat{R}_a^\phi \hat{\rho} (\hat{R}_a^\phi)^\dag$.

We may note that the nonselective post-measurement density operator
(\ref{eq:nonselective}) may be written as
\begin{equation}
    \label{eq:flip}
    \hat{\rho}_a^n = {1 \over 2} \left ( \hat{\rho} +
    \hat{\rho}_a^\pi \right ).
\end{equation}
This is a classical mixture of the initial state and the state where
the phase of the vector corresponding to the measured projector has
been flipped. This is the well known phase-randomization or
decoherence which is associated with nonselective measurements.

It can be shown that the imaginary part of the weak value may be
reconstructed by subjecting a third sub-ensemble to a selective
phase-rotation $\hat{R}_a^{\phi}$. In fact, this works for any phase
angle $\phi$ except $0$ and $\pi$, but the simplest result is
obtained for the angle $\pi/2$, for which we get
\begin{equation}
    \label{eq:imweak}
    \mathrm{Im} B_w(a) = {\mathrm{Tr} \hat{\rho}_a^{\pi \over 2}
    \hat{B} - \mathrm{Tr} \hat{\rho}_a^n \hat{B} \over 2 \,
    \mathrm{Tr} \hat{\rho} \hat{P}_a}.
\end{equation}
This expression suggests a way of reconstructing the imaginary part
of the weak value. It requires two different state preparations. One
where the system is subjected to a nonselective projector
measurement, the other where it is subjected to a selective phase
rotation of $\pi/2$. If there is a difference in the expectation
value of $\hat{B}$ on these two systems, the imaginary part of the
weak value is non-vanishing.

In the particular case where the observable $\hat{B}$ is also a
projector, $\hat{B} = \hat{P}_b$, the imaginary quasi-probability on
the l.h.s. is non-vanishing if and only if an intervening
measurement of $\hat{P}_a$ disturbs the probability of $\hat{P}_b$,
but with respect to the probability of $\hat{P}_b$ on a state that
has been subjected to a selective phase rotation of $\pi/2$. This
shows that there is a close connection between between imaginary
quasi-probabilities and measurement disturbance.

\section{Conclusion}

We have shown that weak values may be reconstructed from a system
subjected to an initial projector measurement followed by a
measurement of the observable. Whereas the initial projector
measurement is a projective measurement in the same manner as the
post-selection in weak measurements, the observable itself can be
measured using any form of interaction. The weak value is
reconstructed by measuring the expectation value of the observable
with and without a preceding projector measurement. This opens the
possibility of verifying the strange predictions of the theory of
weak values without the need for weak measurements. In this way, one
may avoid the the large experimental inaccuracy and the specific
interaction on which weak measurements is grounded. We found that
nonclassical weak values, i.e. a real part exceeding the eigenvalue
range or a non-vanishing imaginary part, are both directly related
to a finite measurement disturbance in this setting.

\section{Acknowledgements}

The author is grateful to Pier A. Mello for useful and interesting
discussions.


\end{document}